\documentclass[twocolumn,showpacs,pra]{revtex4}
\usepackage{graphicx}
\usepackage{amssymb}

\usepackage{dcolumn}
\usepackage{bm}
\usepackage{amsmath}

\begin{document}
\title{Coherence lifetimes of excitations in an atomic condensate due to the thin spectrum}
\author{T. Birol}
\affiliation{Department of Physics, Ko\c{c} University,
Sar{\i}yer, Istanbul, 34450, Turkey}
\author{T. Dereli}
\affiliation{Department of Physics, Ko\c{c} University,
Sar{\i}yer, Istanbul, 34450, Turkey}
\author{\"O. E. M\"ustecapl{\i}o\u{g}lu}
\affiliation{Department of Physics, Ko\c{c} University,
Sar{\i}yer, Istanbul, 34450, Turkey}
\author{L. You}
\affiliation{School of
Physics, Georgia Institute of Technology, Atlanta, Georgia 30332,
USA}
\date{\today}
\begin{abstract}
We study the quantum coherence properties of a finite sized atomic
condensate using a toy-model and the thin spectrum model formalism.
The decoherence time for a condensate in the ground state, nominally
taken as a variational symmetry breaking state, is investigated for
both zero and finite temperatures. We also consider the lifetimes
for Bogoliubov quasi-particle excitations, and contrast them to the
observability window determined by the ground state coherence time.
The lifetimes are shown to exhibit a general characteristic
dependence on the temperature, determined by the thin spectrum
accompanying the spontaneous symmetry breaking ground state.
\end{abstract}

\maketitle

\section{INTRODUCTION}
Although the first observation of superfluid behavior in helium-4
dates back to 1937 \cite{superfluid}, it was not until 1995 that
superfluid associated with Bose-Einstein condensation (BEC)
\cite{bec_original} was discovered in dilute atomic gases
\cite{bec95}. While the atom-atom interactions are too complicated
to handle in liquid state helium-4, a dilute atomic gas opens the
possibility to construct a microscopic theory for superfluidity.
Since 1995, atomic quantum gases have served as excellent fertile
ground for studying quantum coherence properties of matter and for
testing interesting many body theories.

The theoretical and experimental studies of atomic condensates have
also focused on their quantum coherence properties. Shortly after
the initial discovery of BEC, it was understood that a finite sized
condensate, in addition to the usual decoherence due to imperfect
isolation from the environment, suffers from quantum phase diffusion
\cite{walls,you96}, an interaction driven decoherence due to atomic
number fluctuations from within the condensate \cite{hansch}. This
study suggests a third source of decoherence, which we show limits
the lifetime of a quasi-particle excitation from a condensate, based
on the mechanism of thin spectrum as recently proposed and applied
to any quantum system with a spontaneously broken symmetry
\cite{wezel06, wezel05, wezel07}. Our work therefore constitutes a
natural application of the thin spectrum formalism to the highly
successful mean field theory for atomic condensates, where the
condensate is treated as a U(1) gauge symmetry breaking field.

This paper is organized as follows: we begin with a review of a toy
model calculation for the lifetime of the coherent condensate ground
state as well as a squeezed ground state and a thermal coherent
state. We then review the concept of \textit{thin spectrum} and show
how it is connected with spontaneous symmetry breaking and
decoherence of an atomic condensate. In sec. IV, we show how the
quasi-particle excitations of an atomic Bose-Einstein condensate are
affected by the thin spectrum associated with the ground state
condensate. Finally, in sec. V we show how to generalize the idea of
thin spectrum to systems with multiple broken symmetries. Concluding
remarks are provided in sec. VI.

\section{a toy model}
The basic idea of dephasing from the ground state phase collapse
can be understood based on the 'zero-mode' dynamics of a toy model
\cite{toy1,toy2}. The ground state of an $N$ boson system is with
all $N$ bosons in the lowest energy eigenstate, the zero momentum
state for a homogeneous gas. However, it cannot simply be a Fock
state since Bose-Einstein condensation entails a definite phase
from the broken phase U(1) symmetry, while a number state has no
definite phase. A reasonable approximation is to consider a
coherent state occupation for the zero-mode with an amplitude
$\alpha = \sqrt{N}$. Taking $\alpha$ real is equivalent to
explicitly picking a phase of the U(1) symmetry. Because such a
coherent state is not an energy eigenstate, it suffers
phase collapse \cite{walls,you96}. In this section, we discuss the
dynamics of the associated phase collapse based on a simple toy
model to calculate the rate of this collapse and study the
modifications arising from a squeezed ground state.

\subsection{The lifetime for the coherent ground state}
We discuss the zero mode due to BEC, which results in the
breaking the $U(1)$ gauge symmetric
Hamiltonian
\begin{eqnarray}
\mathcal{H}=\frac{\tilde{u}}{2}\hat{a}^\dagger \hat{a}^\dagger
\hat{a} \hat{a} - \mu \hat{a}^\dagger \hat{a},
\label{u1hamiltonian}
\end{eqnarray}
where $\hat{a}$ denotes the atomic annihilation operator for the
condensate (zero) mode. $\tilde{u}$ scales as $\tilde{u}=u_0/V$ with
$V$ the quantization volume and $u_0$ the effective interaction
constant defined as $u_0=4 \pi a_s \hbar^2 /M$. $a_s$ is the s-wave
scattering length and $M$ is the atomic mass. $\mu$ is the chemical
potential, a Lagrange multiplier for fixing the density of the average
number of condensed particles $N$ in the quantization volume $V$. We
consider a variational, symmetry breaking ground state, a coherent
state satisfying $\hat{a} |z\rangle = z |z\rangle$. Such a state can
be formally generated by the displacement operator $D(z)=\exp(z
\hat{a} -z^* \hat{a}^\dagger)$ acting on the vacuum state, or
$D(z)|0\rangle=|z\rangle$. Minimization of the mean free energy
$\langle z| \mathcal{H}|z\rangle$ then fixes $|z|\simeq\sqrt{N}$.
This coherent state can be expanded in terms of the eigenvectors of
the Hamiltonian, e.g. the Fock number states $|n\rangle$ so that
\begin{eqnarray}
|z\rangle = {\rm e}^{-|z|^2/2}\sum_{n=0}^\infty
\frac{z^n}{\sqrt{n!}} |n\rangle .
\end{eqnarray}

In this case the order parameter for BEC, is the
expectation value of the annihilation operator.
In the Heisenberg picture, the operator
$\hat{a}(t)$ is
\begin{eqnarray}
\hat{a}(t)={\rm e}^{\frac{i}{\hbar}\mathcal{H}t}\hat{a}\,
{\rm e}^{-\frac{i}{\hbar}\mathcal{H}t}.
\end{eqnarray}
In terms of the eigenenergy $E_n=\frac{\tilde{u}}{2}(n^2 - n) - \mu n$, defined through
$\mathcal{H}|n\rangle = E_n |n\rangle$, for the $n$-{th} Fock state
$|n\rangle$, one can easily calculate
\begin{eqnarray}
\langle z|\hat{a}|z\rangle
    &=&\sqrt{N} \exp\left(N[{\rm e}^{-\frac{i}{\hbar}
    \tilde{u}t}-1]\right){\rm e}^{\frac{i}{\hbar}\mu t},
\label{gte}
\end{eqnarray}
whose short time behavior is found to be
\begin{eqnarray}
\langle z|\hat{a}|z\rangle = \sqrt{N} {\rm e}^{\frac{i}{\hbar}\mu t}
    {\rm e}^{-i\frac{N\tilde{u}}{\hbar}t} {\rm e}^{-\frac{N \tilde{u}^2}{2
    \hbar^2} t^2},
\end{eqnarray}
i.e., revealing an exponential decay \cite{walls,you96}. At longer
time scale, it turns out that $\langle z|\hat{a}|z\rangle$ revives
due to the discrete, and thus periodic, nature of the exact time evolution
(\ref{gte}).

The short time decay defines a collapse-time proportional to
$t_c \sim \hbar /\sqrt{N}\tilde{u}$.
The ratio of the revival time $t_r$ required for the order parameter
scales as $t_r/t_c =\sqrt{N}$, and becomes
infinite in the thermodynamic limit. In order to get an estimate of
this $t_{c}$, we introduce a characteristic length scale for the
harmonic trap potential as
$a_{\rm ho}=\sqrt{\hbar/(M \omega_{\rm tr})}$, in terms of the
harmonic trap frequency $\omega_{\rm tr}$. Denoting the density of
condensed atom numbers in the quantization
volume as $\rho=N/V$, we find
\begin{eqnarray}
t_c=\frac{\sqrt{N}}{4 \pi N_{\rm eff}}\frac{1}{\omega_{\rm tr}},
\end{eqnarray}
where we have defined $N_{\rm eff}=\rho a_{\rm ho}^2 a_s$.
Assuming a typical situation of current experiments with
$N\sim 10^6$, $a_s=10$ nm, $a_{\rm ho}=1$ $\mu$m, and $\rho=10^{21}$
m$^{-3}$, we get $t_c\simeq 10/\omega_{\rm tr}$. For a magnetic trap
with $\omega_{\rm tr}=100$ Hz, this amounts to $t_c\sim 10^{-1}$
seconds, clearly within the regime to be confirmed and studied
experimentally \cite{hansch}.

\subsection{A squeezed ground state}
The unitary squeezing operator \cite{mandel} for a single bosonic mode
is defined as
\begin{eqnarray}
S(\gamma)={\rm e}^{\frac{\gamma}{2}\hat{a}\hat{a}-
\frac{\gamma^*}{2}\hat{a}^\dagger\hat{a}^\dagger}.
\end{eqnarray}
The squeezed coherent state $|\alpha, \gamma\rangle =
D(\alpha)S(\gamma)|vac\rangle$ is also a minimum uncertainty state,
although its fluctuations in the two orthogonal quadratures are not
generally equal to each other. Fluctuations of one quadrature are
reduced or squeezed at the expense of the other. The arguments of
$\gamma$ and $\alpha$ determine which quadrature is squeezed. In
particular, if both $\gamma$ and $\alpha$ are real, then the state
is a number squeezed state, with the uncertainty in atom number
reduced at the cost of higher uncertainty in the conjugate phase
variable. We expect such a state to have a longer life time, since
the phase collapse speed is generally proportional to $\Delta N$,
which is smaller in this case, as have recently observed
experimentally \cite{mara,ibloch}. A wide phase distribution, on the
other hand, makes the squeezed state more similar to a Fock state
which has a uniform phase distribution, and is less influenced by
the decoherence effect due to the U(1) symmetry breaking field
because of the reduced number fluctuations.

\begin{figure}[h]
\centering{\vspace{0.5cm}}
\includegraphics[width=3.5in]{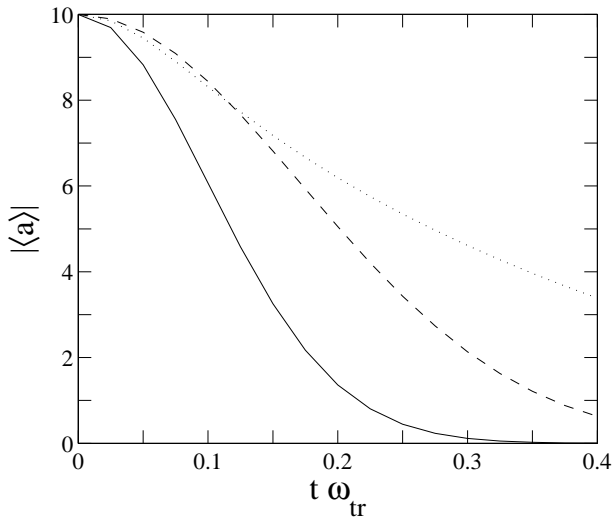}
\caption{The comparison of the short time decay character for a
coherent state condensate with that for a squeezed state at
$\zeta=0.5$ and $\zeta=0.9$. The parameters used are $a_s=10$ nm,
$a_{\rm ho}=1$ $\mu$m, $n=10^{21}$ m$^{-3}$, but now for $N=100$. In
this case, the dimensionless time in units of $\hbar/\tilde{u}$
becomes $\hbar/\tilde{u}= \omega_{\rm tr}^{-1}$. The fastest decay
(solid line) denotes the result for a coherent state, while the
dashed (dotted) line refers to that of a squeezed state with
$\zeta=0.5$ ($\zeta=0.9$). As expected, the choice of a squeezed
state with real parameters $\alpha$ and $\zeta$ improves the
coherence time.} \label{fig1}
\end{figure}

\begin{figure}[h]
\centering{\vspace{0.5cm}}
\includegraphics[width=3.5in]{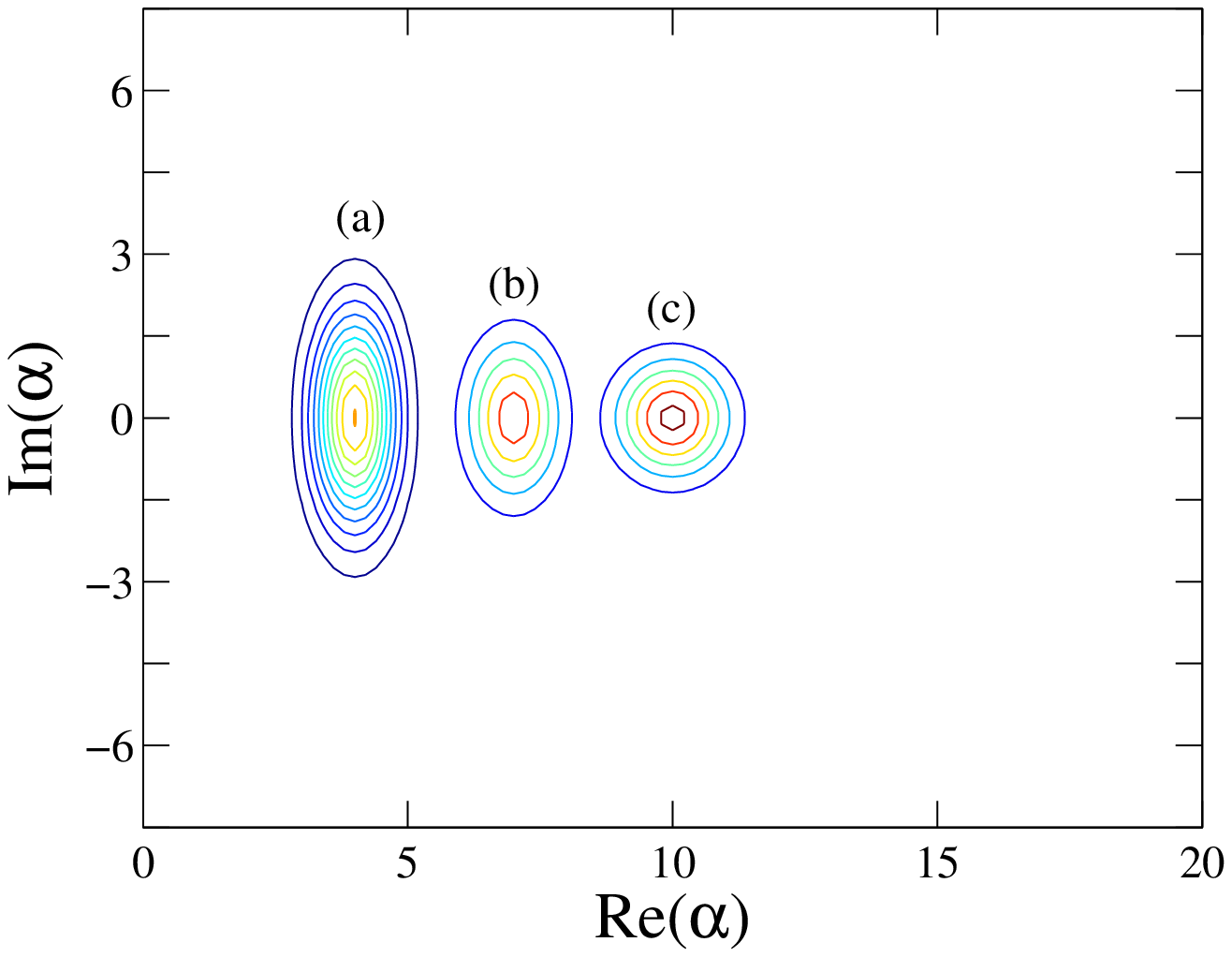}
\caption{(Color online) The phase space distributions of the initial
states used in Fig. \ref{fig1}. Curves (a), (b) and (c) correspond
to $\zeta=0.9$, $\zeta=0.5$ and $\zeta=0$, respectively. Although
all of them should be centered at $\alpha=10$, they are shifted for
convenience. } \label{fig2}
\end{figure}

\begin{figure}[h]
\centering{\vspace{0.5cm}}
\includegraphics[width=3.5in]{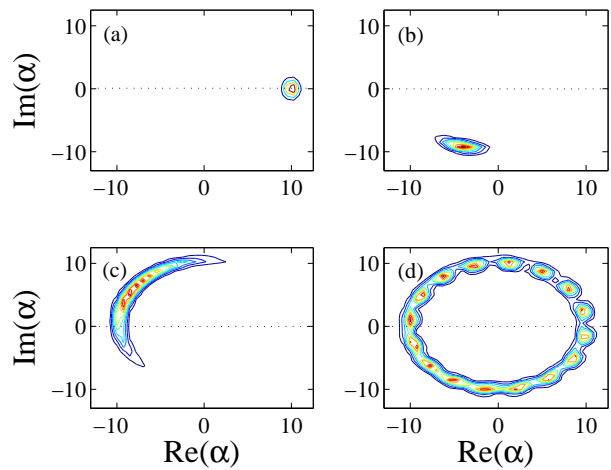}
\caption{(Color online) Time evolution of the Q-function for a
squeezed-coherent state with $\alpha=10$ and $\zeta=0.5$ for
different values of $t \omega_{tr}$. Figures (a), (b), (c) and (d)
shows the Q-function distributions for $t \omega_{tr}=0$, $t
\omega_{tr}=0.02$, $t \omega_{tr}=0.10$ and $t \omega_{tr}=0.40$. It
is  seen that as the order parameter decays, the broken phase
symmetry is restored, since the Q-function distribution becomes
rotationally symmetric. } \label{fig3}
\end{figure}

In order to understand the essence of the above discussion,
we choose to follow similar
arguments as with the coherent state considered previously.
We will study the time evolution
of the single mode state $|\alpha, \gamma\rangle$
subject to the same $U(1)$ gauge symmetric Hamiltonian
(\ref{u1hamiltonian}). For notational convenience we define
\begin{eqnarray}
\zeta= \gamma \frac{\tanh(|\gamma|)}{|\gamma|} .
\end{eqnarray}
The Fock state expansion of the squeezed state in terms of this new
variable is \cite{wunsche}
\begin{eqnarray}
|\alpha, \gamma \rangle
    &=& \sum_{n=0}^{\infty} A_n(\alpha, \zeta) |n\rangle\nonumber\\
    &=& (1-|\zeta|^2)^{1/4}
    {\rm e}^{- \frac{(\alpha + \zeta \alpha^*)\alpha^*}{2}}
    \nonumber\\&&
    \sum_{n=0}^{\infty}
    \sqrt{\frac{\zeta^n}{2^n n!}} H_n\left(\frac{\alpha+\zeta \alpha^*}{\sqrt{2 \zeta}}\right)
    |n\rangle,
\end{eqnarray}
where $H_n$ is the n-th order Hermite polynomial. In the limit
$x\rightarrow\infty$, $H_n(x)$ behaves like $2^n x^n$.
Hence, the squeezed state approaches a coherent
state when $\zeta\rightarrow0$.
The corresponding expectation value for $\hat{a}(t)$
now takes the form
\begin{eqnarray}
\langle \alpha, \gamma | \hat{a}(t) | \alpha,\gamma \rangle =
\sum_{n=0}^{\infty}\sqrt{n+1}A_n^* A_{n+1}
 {\rm e}^{\frac{i}{\hbar}(E_n - E_{n+1})t},
 \label{esg}
\end{eqnarray}
where the complex nature of $A_n(\alpha,\zeta)$ makes the analytic
evaluation of this expression nontrivial. We therefore resort to
numerical studies. In a recent paper, number squeezing of the
initial state by a factor of $10$ was reported \cite{mara}. This
corresponds to $\gamma=\ln{10}$ or $\zeta\simeq 0.98$. In our
numerical calculations, we consider the time evolution of
(\ref{esg}) for $\alpha=10$ at $\zeta=0.5$ and $0.9$. The results in
Fig. \ref{fig1} manifest that squeezing in the particle number
fluctuations improves the coherence time for the condensate. The
phase space distributions of the initial states used in Fig.
\ref{fig1} are displayed in Fig. \ref{fig2}. The longest lived
preparation of the condensate is the one with strongest squeezing in
the particle number or the one with largest phase fluctuations. In
order to examine how the phase distribution evolves in time, we can
look the propagation of the Q-function. We find that all coherent
preparations of the condensate eventually lose the imprinted phase
information, and the system recovers its uniform phase distribution
as in a Fock state. A typical result of our simulations is presented
in Fig. \ref{fig3}. By initially preparing the condensate in a
number squeezed coherent state, with already broad phase
distribution, longer life times of the condensate are achieved.

\begin{figure}[h]
\centering{\vspace{0.5cm}}
\includegraphics[width=3.5in]{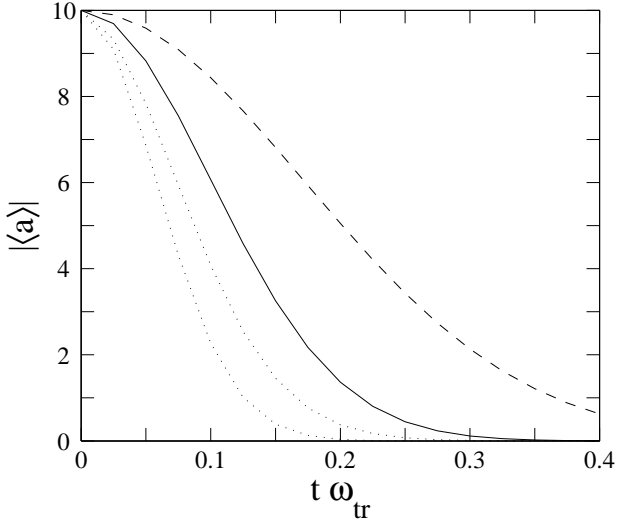}
\caption{Decay of the order parameter for the coherent state and
squeezed states of $\zeta=0.5$, $\zeta=0.5 i$ and $\zeta=-0.5$ as a
function of $t \omega_{tr}$. The solid line is the coherent state,
the dashed line is the squeezed state with $\zeta=0.5$, and the
dotted ones are the squeezed states with $\zeta=0.5i$ and
$\zeta=-0.5$. Only the state with real squeezing parameter has a
longer life time than the coherent state.} \label{fig4}
\end{figure}

More generally, our discussions can reach beyond the choices of real
parameters $\alpha$ and $\zeta$. Consequently, different results may
be expected, as we illustrate the comparisons between a coherent
state and squeezed states with $\zeta=0.5$, $\zeta=0.5 i$, and
$\zeta=-0.5$ in Fig. \ref{fig4}. We see that the last two choices of
the squeezing parameters lead to reduced coherence times, a result
that again can be reasonably understood in terms of the increased
uncertainty in the atom number, as it causes faster collapse.

\begin{figure}[h]
\centering{\vspace{0.5cm}}
\includegraphics[width=3.5in]{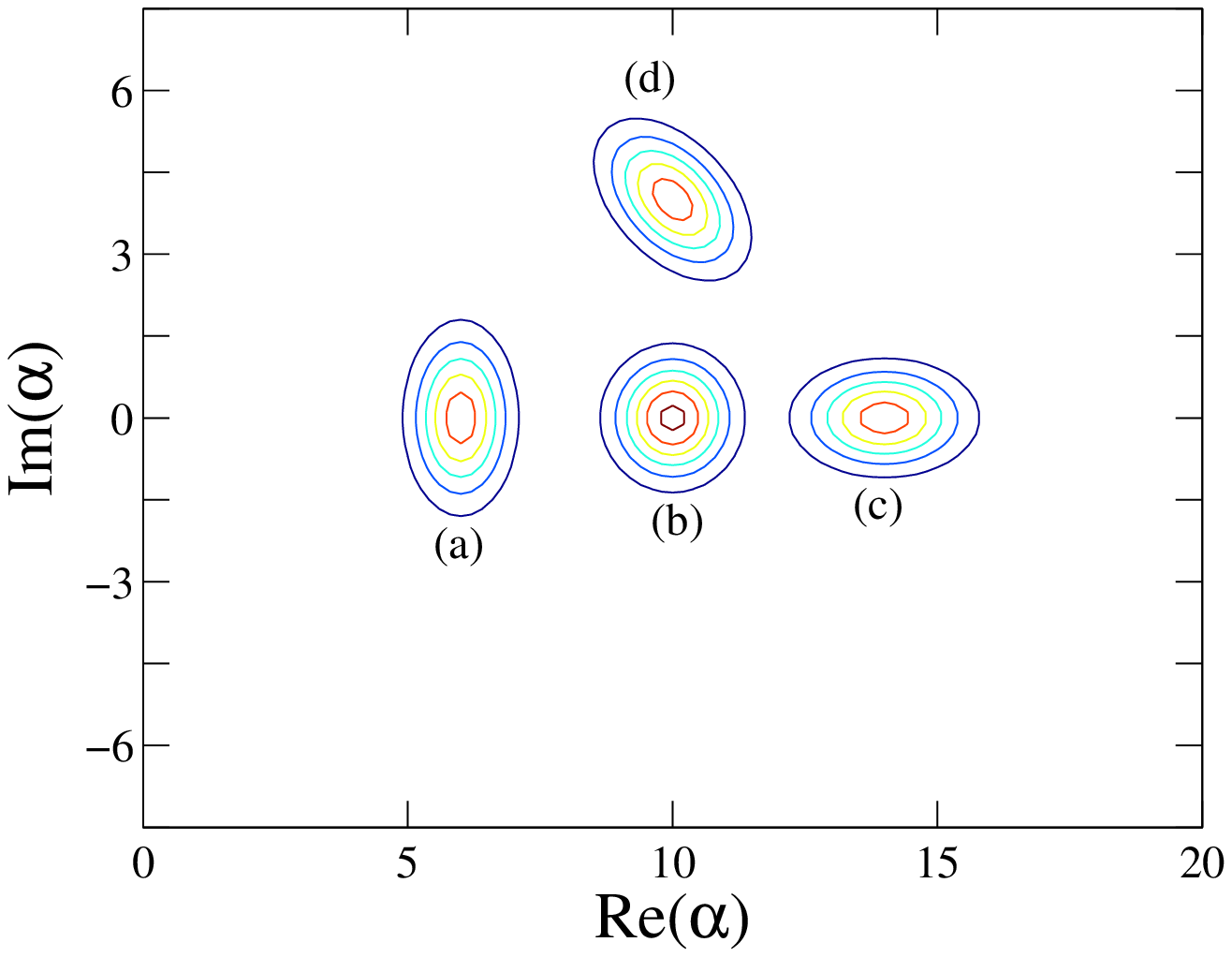}
\caption{(Color online) The phase space distributions of the initial
states used in Fig. \ref{fig4}. Curves (a), (b), (c) and (d)
correspond to $\zeta=0.5$, $\zeta=0$, $\zeta=-0.5$ and $\zeta=0.5
i$, respectively. Although all of them should be centered at
$\alpha=10$, they are shifted for convenience. } \label{fig5}
\end{figure}

\subsection{A thermal coherent state}
\begin{figure}[h]
\centering{\vspace{0.5cm}}
\includegraphics[width=3.5in]{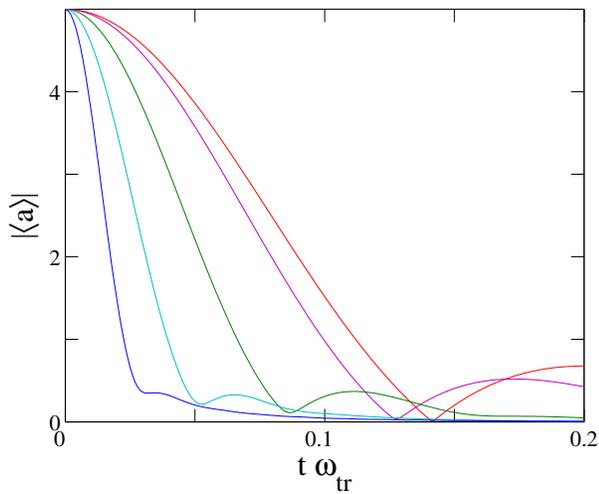}
\caption{(Color online) The short time decays for thermal coherent
states. The lines correspond respectively to $T=1000$ nK, 100 nK, 10
nK, 1 nK, and 0.001 nK from left to right. The humps are entirely
due to the ground degeneracy $E_0=E_1$. Even as the temperature
approaches zero, the state (\ref{eqt2}) does not approach the
ordinary coherent state $D(\alpha) |0\rangle$. Instead, it is a
superposition state $D(\alpha) (|0\rangle + |1\rangle)/\sqrt{2}$.}
\label{fig6}
\end{figure}

To extend the above discussions to finite temperature systems,
we will now introduce the thermal coherent state, which
possesses both a thermal character as well as a phase.
Consider the following density matrix for a thermal state
\begin{eqnarray}
\rho_{\rm th}   &=& {\rm e}^{-\beta \mathcal{H}}\nonumber\\
            &=& \sum_n {\rm e}^{- \beta E_n} |n\rangle\langle n|,
\label{eqt}
\end{eqnarray}
where $\beta = 1/k_B T$, $k_B$ is the Boltzmann constant and $T$ is
the temperature. In this state (\ref{eqt}), $\mathcal{H}$ is the
Hamiltonian operator, $\mathcal{H}=\tilde{u} \hat{a}^\dagger
\hat{a}^\dagger \hat{a} \hat{a} /2$. $E_n$ is redefined,
corresponding to $\mathcal{H}|n\rangle = E_n |n\rangle$. $\rho_{\rm
th}$ is a mixed state that has a thermal character but not a
definite phase. In order to introduce a coherent component, and also
to change the mean number of atoms, we can make use of the
displacement operator
\begin{eqnarray}
    \rho = D(\alpha) \rho_{\rm th} D^\dagger (\alpha).
\label{eqt2}
\end{eqnarray}
We shall call this state a thermal coherent state,
whose properties can be conveniently studied with the
aid of the generalized
coherent or displaced number states \cite{roy82}
\begin{eqnarray}
    D(\alpha)|n\rangle  &=& |n, \alpha\rangle \nonumber\\
                        &=& \sum_{n=0}^{\infty} {\rm e}^{- \frac{1}{2}
                        |\alpha|^2} \sqrt{\frac{n!}{m!}}
                        \alpha^{m-n} \text{L}_n^{m-n}(|\alpha|^2)|m\rangle \hskip 24pt\nonumber\\
                        &=& \sum_{n=0}^{\infty} C_m (n, \alpha)
                        |m\rangle,
\end{eqnarray}
where $L_k^l$ is the generalized Laguerre Polynomial.
The thermal coherent density matrix now becomes
\begin{eqnarray}
    \rho    &=& \sum_{n=0}^\infty {\rm e}^{-\beta E_n} D(\alpha)
            |n\rangle\langle n|D^\dagger(\alpha)\nonumber\\
            &=& \sum_{n=0}^\infty {\rm e}^{-\beta E_n} |n,
            \alpha\rangle \langle n,\alpha|\nonumber\\
            &=&\sum_{n m m'} {\rm e}^{-\beta E_n} C_m(n,\alpha)
            C_{m'}^*(n,\alpha) |m\rangle \langle m'|,
\end{eqnarray}
with which we can again consider the time evolution of the
expectation value of $\hat{a}(t)$,
\begin{eqnarray}
\langle \hat{a}(t) \rangle &=& \sum_{nmm'k} {\rm e}^{-\beta E_n}
C_m(n,\alpha)
    C_{m'}^\dagger(n,\alpha) \langle k|m\rangle \nonumber\\
    &&\langle m'|
    {\rm e}^{\frac{i}{\hbar}\mathcal{H}t}\hat{a}
    {\rm e}^{-\frac{i}{\hbar}\mathcal{H}t}|k\rangle,
\end{eqnarray}
calculated according to
$\langle \hat{a}(t) \rangle = \text{Tr}\left( \rho \hat{a}(t) \right)$.
In the end, we find
\begin{eqnarray}
\langle \hat{a}(t) \rangle &=& \sum_{n m} {\rm e}^{-\beta E_n}
C_{m+1}(n, \alpha) C_{m}^*(n, \alpha) \nonumber\\
&&\sqrt{m}\,
{\rm e}^{-\frac{i}{\hbar}(E_{m+1}-E_m) t}.
\end{eqnarray}
In general, the phase factors will interfere destructively in the
above. The thermal distribution weight ${\rm e}^{-\beta E_n}$ term
determines how many different terms contribute. This implies that
the temperature definitely leads to a reduced coherence time for the
state. For the initial preparations of a condensate of $100$ atoms,
as depicted in Fig. \ref{fig5}, Fig. \ref{fig6} illustrates the
decay of these condensates at various temperatures.

\section{The thin spectrum formalism}
\subsection{A simple theory}
By the thin spectrum, we typically refer to a group of states whose
energy spacings are so low that they are not exactly controllable in
any experiments. The effect of such states on the partition function
and on the decoherence has been  studied extensively, for instance
see Ref. \cite{wezel06}. In many-body systems, models of thin
spectra arise quite often whenever there exists a spectrum with
level spacing inversely proportional to the system size. These
states with vanishing energy difference in the thermodynamic limit
are usually beyond experimental reach and therefore constitute a
thin spectrum.

We begin by reviewing the ideas developed in \cite{wezel06},
which use two quantum numbers: $n$ and $m$, to denote the thin spectrum and
ordinary states. When the initial state is prepared at $m=0$, the
thin spectrum distribution will be a thermal one.
This leads to the initial state for the system being
\begin{eqnarray}
\rho({t=0})=Z^{-1}\sum_n {\rm e}^{-\beta E_0^{(n)}} |0,n\rangle \langle
0,n|,
\end{eqnarray}
where $\mathcal{H}|m,n\rangle = E_m^{(n)} |m,n\rangle$.
$Z$ is the partition function, $Z=\sum_n \exp{(-\beta E_0^{(n)})}$.
A transformation $|0,n\rangle \rightarrow \sum_m C_m |m,n\rangle$,
leads to the state
\begin{eqnarray}
\rho=Z^{-1}\sum_{nmm'} {\rm e}^{-\beta E_0^{(n)}} C_m
C_{m'}^*|m,n\rangle \langle m',n|,
\end{eqnarray}
which becomes
\begin{eqnarray}
\rho({t>0}) &=&\sum_{nmm'} \frac{ {\rm e}^{-\beta E_0^{(n)}}}{Z}
{\rm e}^{-\frac{i}{\hbar}(E_m^{(n)} -E_{m'}^{(n)})t}\nonumber\\
&&C_m C_{m'}^*|m,n\rangle
\langle m',n|,
\end{eqnarray}
after time evolution.
When it is observed, the details of this density matrix
cannot be seen, since the thin spectrum is assumed to be
beyond experimental reach.
Therefore, only the reduced density matrix, which is obtained
by taking the trace of $\rho$ over the thin spectrum states, is
observed. Following the work of \cite{wezel06}, we define the thin
spectrum state
$|j_{\rm thin}\rangle$ by $\langle j_{\rm thin}|m,n\rangle= \delta_{j,n}
|m\rangle$ where $|m\rangle$ denotes the ordinary observable state of
a system. This then allows us to compute the reduced system state
\begin{eqnarray}
\rho^{(\rm red)}
    &=& \sum_j \langle j_{\rm thin}|\rho({t>0})|j_{\rm thin}\rangle\nonumber\\
    &=&\sum_{mm'n}\frac{ {\rm e}^{-\beta E_0^{(n)}}}{Z} {\rm e}^{-
    \frac{i}{\hbar} (E_m^{(n)} -E_{m'}^{(n)})t}C_m
    C_{m'}^*|m\rangle \langle m'|.\hskip 24pt
\label{equ_collapse}
\end{eqnarray}
While the diagonal elements $\rho^{(\rm red)}_{m m}=|C_m|^2$
experience no time evolution, the off diagonal elements suffer a
phase collapse unless $E_m^{(n)} -E_{m'}^{(n)}$ is independent of
$n$. For a two state system $(m=0,1)$, the off-diagonal element will
decay at a rate $\Delta E_{\rm thin}/E_{\rm thin}$ with $\Delta
E_{\rm thin}=E_1^{(n)} -E_0^{(n)}$ and $E_{\rm thin}=E_0^{(n)}$
\cite{wezel06}.

\subsection{Continuous symmetry breaking and the Goldstone theorem}
The Nambu-Goldstone Theorem \cite{goldstone} dictates the existence
of a gapless mode whenever a continuous symmetry is broken
spontaneously. For a ferromagnetic material, this mode is the long
wavelength spin waves \cite{ezawa_chp4}. For a crystalline structure,
when the translational symmetry is broken, the Nambu-Goldstone mode
(NGM) describes the overall motion of the crystal \cite{wezel06}. For an
atomic condensate, where the BEC leads to the breaking of the gauge
symmetry, the corresponding gapless mode induces phase displacement
of the condensate \cite{you96,forster166}.

Consider a diagonal Hamiltonian, which may
correspond to normal mode excitations with different $\omega$'s,
\begin{eqnarray}
\mathcal{H}=\sum_k \hbar \omega_k b^\dagger_k b_k,
\end{eqnarray}
where $b_k$ is the annihilation operator for the $k$-{th} mode.
As usual, the bosonic commutation relations are
assumed, $[b_{k'},b_k^\dagger]=\delta_{k,k'}$ and
$[b_k,b_{k'}]=[b_k^\dagger,b_{k'}^\dagger]=0$. If there is a broken
symmetry, motion along the axis of this symmetry will experience no
restoring force, and hence the Hamiltonian of this mode will have the
form $p^2 /2I$ rather than $a^\dagger a$, where $p$ is the
corresponding momentum operator and $I$ is the corresponding
inertia mass. Hence, the Hamiltonian becomes
\begin{eqnarray}
\mathcal{H}=\frac{1}{2 I}  p^2 + \sum_k \hbar \omega_k b^\dagger_k
b_k . \label{equ_sampleH}
\end{eqnarray}
The Hamiltonians for both a crystal \cite{wezel06} and a condensate
\cite{you96} can be shown to take this form. In both cases, the
inertia mass parameter $I$ depends on the total atom number $N$ and
can either diverge or vanish in the thermodynamic limit
when $N\rightarrow \infty$.

The relationship between the Nambu-Goldstone Theorem and the thin
spectrum is that the NGM guarantees the existence of a gapless mode,
with the corresponding momentum $p$ taking an arbitrarily small
value. Therefore, the value of $p$ is always capable of giving rise
to thermal fluctuations below the experimental precision and every
NGM corresponds to a thin spectrum \cite{wezel06}.

\subsection{An explicit calculation}
\label{subsext_explicit}

\begin{figure}[h]
\centering{\vspace{0.5cm}}
\includegraphics[width=3.5in]{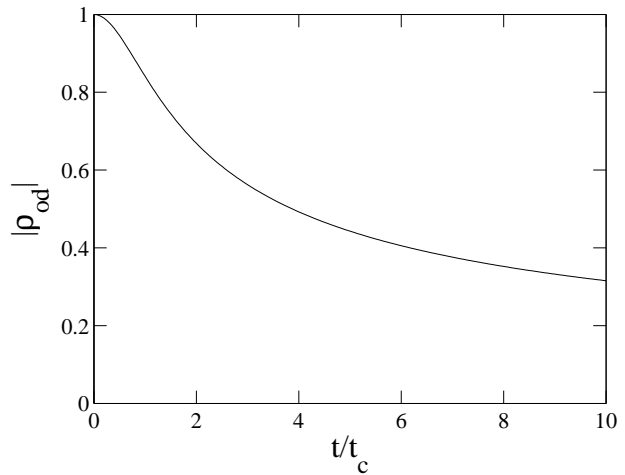}
\caption{The decay of $|\rho_{\rm od}^{(\rm red)}|$ as a function of
$t/t_{c}$.} \label{fig7}
\end{figure}

The Hamiltonian (\ref{equ_sampleH}) is very common, Therefore, it is
useful and instructive to calculate its time of collapse explicitly.
According to the thin spectrum theory, the general state of a system
takes the form $|p,\{N_k\}\rangle$ denoted by two sets of quantum
numbers $p$ and $\{N_k\}$. For simplicity, we assume that both $p$
and $k$ are one dimensional quantities. Furthermore, only two
different states of the system are considered in order to use it as
a qubit. Assume that the elementary excitation which brings the
system from $\{N_k\}$ to $\{N'_k\}$ has a corresponding energy
$\epsilon$. In general, such an excitation may also change the
inertia mass $I$ of the $p$ term. For example, an interstitial
excitation changes the total mass of the crystal \cite{wezel06}.
Similarly, an excitation inside an atomic condensate can change its
peak density, which determines the inertia mass factor in front of
the phase coordinate \cite{you96,alpha}. Such a change is necessary
for our mechanism of phase diffusion to occur. When this change to
the effective mass from $I$ to $I(1+\delta)$ is small against the
small change $\delta$ of the parameter, the off-diagonal element in
equation (\ref{equ_collapse}) evolves in time as
\begin{eqnarray}
\rho^{(\rm red)}_{\rm od}= Z^{-1} \left[\sum_p {\rm e}^{-\beta E_0^{(p)}}
{\rm e}^{-\frac{i}{\hbar} (E_1^{(p)} -E_0^{(p)})t}\right] C_1 C_0^*,
\label{2statecollapse}
\end{eqnarray}
where $E_0^{(p)}= p^2/2I$ and $E_1^{(p)}=\epsilon + p^2/2I(1+\delta)$.
Upon substituting into the above, we find
\begin{eqnarray}
\rho^{\rm (red)}_{\rm od}= Z^{-1} {\rm e}^{-\frac{i}{\hbar}
\epsilon}\left[\sum_p
{\rm e}^{-\left(\frac{\beta}{2I}-\frac{i}{2\hbar}\frac{\delta}{I}t\right)p^2
}\right] C_1 C_0^* .
\end{eqnarray}
Since $p$ is continuous, its summation becomes an integral, or
\begin{eqnarray}
\rho^{\rm(red)}_{\rm od}
    &=& Z^{-1} {\rm e}^{-\frac{i}{\hbar}\epsilon}\frac{\sqrt{\pi}}{2}
        \frac{1}{\sqrt{\left(\frac{\beta}{2I}-2\frac{i}{\hbar}\frac{\delta}{I}t\right)}}
        C_1 C_0^*,
\end{eqnarray}
which gives
\begin{eqnarray}
|\rho^{\rm (red)}_{\rm od}|^2=({\rm const.}) \frac{1}{\sqrt{1 + 16 t^2 \delta^2 /
\beta^2 \hbar^2}}.
\end{eqnarray}
Thus, the off diagonal term decays in a time
\begin{eqnarray}
t_{c}\sim\hbar/k_B T \delta,
\label{t_spon_delta}
\end{eqnarray}
as seen in Fig. \ref{fig7}.

To apply the above result to an atomic condensate, we consider the
relevant temperature scale at $T \sim 100 nK$ and assume that
a particular observable excitation has $\delta \sim 10^{-1}$. In this
case, we see that $t_{c}\sim 10^{-3}$ seconds, less than the
life times of many observed ground states.
We can also try to obtain an
approximation to the coherence time of the condensate ground state.
Taking the atom number as $N \sim 10^6$, if the ground state
is assumed a coherent state, than the number fluctuations is of
the order of $\Delta N =\sqrt{N}$. The inertia parameter $I$
is proportional to $I\sim N^{2/5}$ \cite{you96, alpha} in the Thomas-Fermi
limit, which gives
$\delta=\left[(N+\Delta N)^{2/5} -N^{2/5}\right]/N^{2/5}= 2\Delta N
/5N$, or $\delta \sim 10^{-3}$. Substituting this in, we find $t_{c}
\sim 10^{-1}$ seconds, much larger than for the excited state,
as to be expected. Furthermore, the
result for the ground state life time is in agreement with our
previous calculation in sect. II.

\section{quasi-particles in a condensate}
\subsection{A thermal state}
We now focus on the Hamiltonian of a dilute, weakly interacting
atomic Bose gas \cite{huang}
\begin{eqnarray}
\mathcal{H}=\sum_k E_k a_k^\dagger a_k +
\frac{\tilde u}{2}\sum_{k,p,q}a_{p+q}^\dagger a_{k-q}^\dagger a_k a_p.
\end{eqnarray}
Omitting the $3^{rd}$ and $4^{th}$ order operator
terms in the non-condensed mode (${k\neq 0}$), we can
partition the Hamiltonian into
\begin{eqnarray}
\mathcal{H} &=& \mathcal{H}_z + \mathcal{H}_e, \\
\mathcal{H}_z &=& \frac{\tilde u}{2} (\hat{n}_0^2 -\hat{n}_0),\nonumber \\
\mathcal{H}_e &=& \sum_{k\neq0}\left[\left(E_k +2\tilde{u}
            \hat{n}_0\right)\hat{n}_k +
            \frac{\tilde u}{2}\left(a_k^\dagger a_{-k}^\dagger a_0 a_0 +
            h.c.\right)\right],\nonumber
\end{eqnarray}
where $\hat{n}_k=a_k^\dagger a_k$ is the mode occupation.
The above two parts of the
Hamiltonian actually do not commute with each other, as
we can easily check that
\begin{eqnarray}
[\mathcal{H}_z,\mathcal{H}_e]=\frac{\tilde{u}^2}{2}
\sum_{k\neq 0} \left[a_k^\dagger a_{-k}^\dagger\left(-\hat{n}_0
a_0^2 - a_0^2 \hat{n}_0+a_0^2\right) - h.c.\right]. \hskip 12pt
\end{eqnarray}
Neglecting the quantum nature of $a_0$ in $\mathcal{H}_e$, we
can replace $\hat{n_0}/ V$ by $\rho_0=N_0/V$, and get
\begin{eqnarray}
\mathcal{H}_e=\sum_{k \neq 0}\left[\epsilon_k \hat{n}_k + \frac{u_0
\rho_0}{2}(a_k^\dagger a_{-k}^\dagger + h.c.)\right],
\label{Hbg}
\end{eqnarray}
where we have defined $\epsilon_k=E_k + 2 u_0 \rho_0-\mu_0$ with
$\mu_0= u_0\rho_0$ for a coherent condensate state. In this
approximation, $[\mathcal{H}_z,\mathcal{H}_e]=0$ at the cost of
sacrificing the conservation of $N_{\rm total}=\sum_k n_k$.

The quadratic Hamiltonian (\ref{Hbg}) can be diagonalized
with the Bogoliubov quasi-particles into the canonical form
\begin{eqnarray}
\mathcal{H}_e=\sum_{k \neq 0} \omega_k b_k^\dagger b_k + const.,
\end{eqnarray}
with $\omega_k=[\epsilon_k^2 - u_0^2 \rho_0^2]^{1/2}$ \cite{greiner}
and $b_k=S a_k S^{-1}$. $S$ is the multi-mode squeeze operator
\cite{haque06}.

In order to conserve the particle number density in the condensate,
we include a chemical potential term in the zero mode Hamiltonian
\begin{eqnarray}
\mathcal{H}_z=\frac{\tilde u}{2} (\hat{n}_0^2 -\hat{n}_0) -\mu_0
\hat{n}_0.
\end{eqnarray}
The ground state of such a system will be a Fock number state
$|N_0\rangle$ with $N_0 = \mu_0 V /u_0 + 1/2$. We assume that
although $N_0$ and $V$ may fluctuate, their ratio $\rho_0$ is
always a constant, as in the thermodynamic limit.
In this case, $\mathcal{H}_z$ becomes
\begin{eqnarray}
\mathcal{H}_z=\frac{u_0 \rho_0}{2 N_0} (\hat{n}_0^2 -\hat{n}_0)
-\mu_0 \hat{n}_0 .\label{eq:hz}
\end{eqnarray}
Substituting $\mu_0=u_0 \rho_0 -u_0\rho_0/2N_0$, we get
\begin{eqnarray}
\mathcal{H}=\frac{u_0 \rho_0}{2 N_0} \hat{n}_0^2 -\rho_0 u_0
\hat{n}_0 + \mathcal{H}_e .
\end{eqnarray}
For a coherent condensate state with $\mu_0= u_0\rho_0$, we could
immediately get this result by neglecting the second term in Eq.
(\ref{eq:hz}), consistent with the non-zero momentum part of the
Hamiltonian.

The ground state, which we denote as
$|\Phi\rangle$, has $N_0$ bosons in the zero momentum state and no
quasi-particle excitations at all, i.e., for $k\neq0$,
\begin{eqnarray}
b_k |\Phi\rangle  &=& 0 \nonumber \\
                &=& S a_k S^{-1} |\Phi\rangle,
\end{eqnarray}
or
\begin{eqnarray}
a_k S^{-1}|\Phi\rangle=0,
\end{eqnarray}
and
\begin{eqnarray}
|\Phi\rangle=S|{\rm vac}\rangle.
\end{eqnarray}
Therefore, the quasi-particle vacuum state
$|\Phi\rangle$ is in fact a
squeezed vacuum of atoms with nonzero $k$.

Now, we consider a setup with $n$ atoms in the condensate mode
and $m$
quasi-particle excitations at a certain, single $k$ mode,
while all other modes are empty.
We will denote such a state by $|n,m\rangle$
\begin{eqnarray}
\hat{n}_0|n,m\rangle &=& n |n,m\rangle,\\
\hat{n}_{k'}|n,m\rangle &=& m\,\delta_{k,k'} |n,m\rangle .
\end{eqnarray}
In the single-particle excitation regime $E_k \gg \rho_0 u_0$, each
quasi-particle excitation reduces the number of condensate atoms by
one. In this case, the energy of this state can be written as
\begin{eqnarray}
\mathcal{H}|n,m\rangle  &=& E_m^{(n)} |n,m\rangle \nonumber\\
                        &=& \left[\frac{u_0 \rho_0n^2}{2(N_0 - m)} - u_0 \rho_0 n + m
                        \omega\right]|n,m\rangle, \hskip 12pt
\end{eqnarray}
where we simply denote $\omega=\omega_k$.

Assume the system can be initially prepared with no quasi-particle
excitation at all, but is in a Boltzmann weighted distribution
over the states $|n,0\rangle$, i.e.,
\begin{eqnarray}
\rho(t=0)\propto\sum_n {\rm e}^{- \beta E_0^{(n)}} |n,0\rangle\langle
n,0|.
\end{eqnarray}
This state will allow us to study
the number fluctuations due to unknown
nonzero temperature constituents that make up the
occupations of the thin spectrum \cite{wezel06}.
The summation index can take any positive integers
and therefore the
summation should be over $0 \leq n < +\infty$.
However, we note that the maximum of $E_0^{(n)}$
is at $N_0 \gg 1$, and because it becomes
extremely small for small values of $n$, we can extend
the summation to be over the full range
$-\infty < n < +\infty$ and replace it with an
integral in the continuous limit as done in the following.

\begin{figure}[h]
\centering{\vspace{0.5cm}}
\includegraphics[width=3.5in]{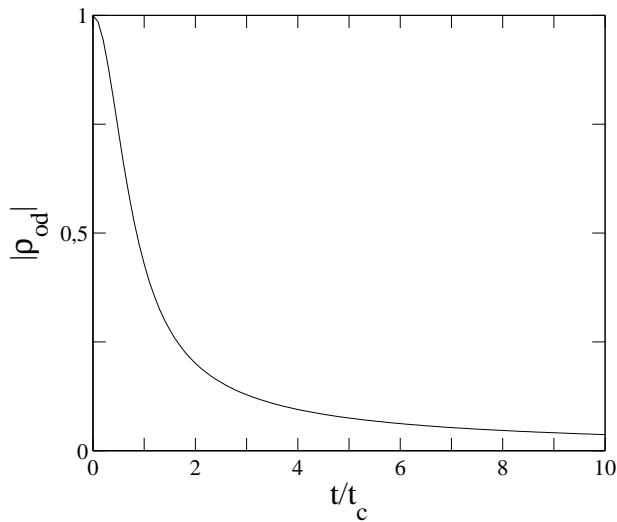}
\caption{The relative decay of the off diagonal element in equation
(\ref{decay_4}) as a function of $t/t_{c}$ for unit values of
parameters. } \label{fig8}
\end{figure}

Excitation of a quasi-particle brings each $|n,0\rangle$ to
$|n,1\rangle$. The off diagonal element of the resulting state
will evolve according to
\begin{eqnarray}
\rho_{\rm od}(t>0) &\propto& \int_{-\infty}^{\infty}  {\rm e}^{-\beta
E_0^{(n)}}
{\rm e}^{-\frac{i}{\hbar}(E_1^{(n)} - E_0^{(n)}) t} dn \nonumber\\
&\propto& \int_{-\infty}^{\infty} {\rm e}^{(-\beta u_0
                \rho_0/2 N_0 + i t u_0 \rho_0/2\hbar N_0^2)n^2 +
                \beta \rho_0 u_0 n} dn \nonumber\\
                &\propto& \sqrt{\pi} \frac{\exp\left(\frac{\beta^2
                \rho_0^2 u_0^2}{2\beta u_0 \rho_0 / N_0 - 2 i t u_0
                \rho_0 / \hbar N_0^2}\right)}{\sqrt{\beta u_0 \rho_0 / 2N_0 - i t u_0 \rho_0 / 2 \hbar
                N_0^2}},
\end{eqnarray}
which gives
\begin{eqnarray}
|\rho_{\rm od}(t)|^2 \propto \frac{\exp\left(\frac{\beta^3 N_0^3
u_0 \rho_0}{\beta^2 N_0^2 + t^2/\hbar^2} \right)}{\sqrt{\beta^2 +
t^2 /\hbar^2 N_0^2}},
\label{decay_4}
\end{eqnarray}
after omitting terms with only a phase factor. Although the
denominator and the numerator have quite different forms, we find
that both decay in a time proportional to $t_{c}\sim \hbar N_0 / k_B
T$. This is the same result that Wezel \textit{et. al.} have found
for a crystal \cite{wezel06}. The decay of this function is plotted
in Fig. \ref{fig8} for unit values of parameters.

For an atomic Bose-Einstein condensate, the relevant parameters are
$N_0\sim 10^6 - 10^8$ and $T\sim 10^{-8} - 10^{-7}$ K. These then
lead to $t_{c} \sim 10^2 - 10^5$ seconds, which is a time much larger
than both the theoretical and observed ground state life times. However,
this is the life time for a single quasi-particle excitation, i.e.,
for $m=1$. It is easy to show that the collapse time is inversely
proportional to $m$ for m values not too large. An easily tractable
excitation should have $m \sim N_0$ and this gives $t_{c} \sim
10^{-4} - 10^{-3}$ seconds, much smaller than both the observed
and expected ground state life times.

The study of temperature dependence for the damping rates of
Bogoliubov excitations of any energy has been carried out before
using perturbation theory. A linear temperature dependence was
found \cite{gora}, surprisingly coinciding with the linear
dependence found here based on the decoherence of the thin spectrum.
Our result clearly would make a quantitative contribution to the
total decay of the quasiparticles, although we note that our
calculation is limited only to the single-particle excitation regime
as we have used $\epsilon_k=E_k\gg u_0\rho_0$. In the phonon branch
corresponding to the low-lying collective excitations out of a
condensate, more complicated temperature dependencies may occur
\cite{liu}. In contrast to damping mechanisms based upon excitation
collision processes in the condensate, the thin spectrum caused
decay rate shows no system specific dependencies, apart from the
dependencies on temperature and the number of atoms. It is
independent of the interatomic interaction strength or the
scattering length, and the quasiparticle
spectrum. This is due to the fact that thin spectrum emerges as a
result of a global symmetry breaking in a quantum system so that
local properties of the system do not contribute to the associated
decay rate.

\subsection{A thermal coherent state}
We now generalize the above idea to a thermal coherent occupation of the
zero-mode. The initial density matrix becomes in this case
\begin{eqnarray}
\rho_{\rm od}(0) &=& Z^{-1} \sum_n {\rm e}^{-\beta E_0^{(n)}}
                    D(\alpha)|n,0\rangle
                    \langle n,0|D^\dagger(\alpha)\nonumber\\
                    &=& Z^{-1} \sum_{n m m'}{\rm e}^{-\beta E_0^{(n)}} C_m(n, \alpha) C_{m'}^*(n, \alpha)\nonumber \\
          &&    |m, 0\rangle \langle m', 0| .
\end{eqnarray}

The system is now brought into a superposition of no quasi-particle
and one quasi-particle state, i.e., $|n, 0\rangle \rightarrow (|n,
0\rangle + |n, 1\rangle)/\sqrt{2}$. After further time evolution,
the state becomes
\begin{eqnarray}
\rho(t) &=&Z^{-1} \sum_{nmm'} \sum_{k k'=0, 1} \frac{{\rm e}^{- \beta
                    E_0^{(n)}}}{2} C_m(n, \alpha) C_{m'}^* (n, \alpha)
                    \nonumber\\
                    &&{\rm e}^{-\frac{i}{\hbar}(E_1^{(m)} - E_0^{(m')})t}
                    |m, k\rangle \langle m', k'|,
\end{eqnarray}
giving rise to the
reduced density matrix and its off diagonal element below
\begin{eqnarray}
\rho^{\rm (red)} &=& Z^{-1} \sum_{nl} \sum_{k k'=0,1} \frac{{\rm e}^{-\beta
                    E_0^{(n)}}}{2}|C_l(n, \alpha)|^2 \nonumber\\
                    &&{\rm e}^{-\frac{i}{\hbar}(E_k^{(l)} - E_{k'}^{(l)})t}
                    |k\rangle \langle k'|,\\
\rho^{\rm (red)}_{\rm od} &=& Z^{-1} \sum_{nl} \frac{{\rm e}^{-\beta E_0^{(n)}} |C_l(n,
\alpha)|^2}{2} {\rm e}^{- \frac{i}{\hbar} (E_1^{(l)} - E_0^{(l)}) t}. \hskip 24pt
\end{eqnarray}

\begin{figure}[h]
\centering{\vspace{0.5cm}}
\includegraphics[width=3.5in]{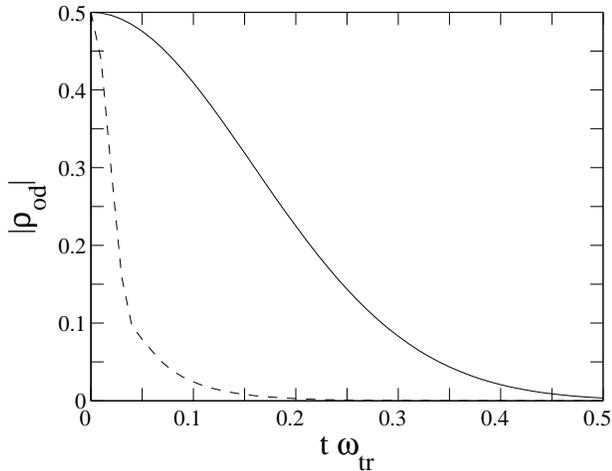}
\caption{Decay of the off diagonal element at $T=10$ nK as a
function of $t \omega_{tr}$. Dashed line shows the decay in the case
of the thermal coherent occupation and the solid line shows that in
the case of the thermal occupation.} \label{fig9}
\end{figure}
\begin{figure}[h]
\centering{\vspace{0.5cm}}
\includegraphics[width=3.5in]{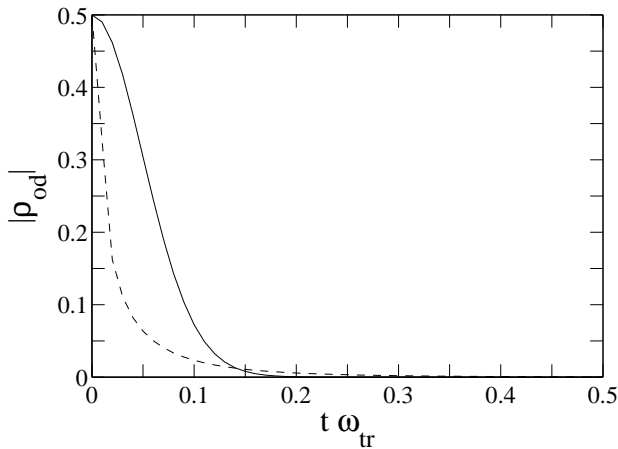}
\caption{Decay of the off diagonal element for $T=100$ nK and
thermal coherent occupation of the zero mode as a function of $t
\omega_{tr}$. Dashed line shows the decay in the case of the thermal
coherent occupation and the solid line shows that in the case of the
thermal occupation.} \label{fig10}
\end{figure}

Figs. \ref{fig9} and \ref{fig10} show the early time decay at
temperatures of $10$ nK and $100$ nK respectively. It is seen that
the decay time for a thermal occupation, which we have studied in
the preceding subsection, exhibits stronger temperature sensitivity,
whereas the decay time for the thermal coherent occupation is not
changed very much by temperature. Therefore, we conclude that for a
thermal coherent occupation, the main reason for the decay of the
off diagonal element is the decay of the zero mode distribution.
However, if there is solely thermal occupation, no decay of the zero
mode occurs, and the off diagonal element decays only because of the
temperature. In the previous section, we have seen that the decay rate
due to the thin spectrum and the decay rate due to the excitation
collision processes show the same temperature dependence
qualitatively. In the case of coherent thermal occupation of the
zero mode, qualitative differences appear in the temperature
dependence of the decay time due to the different decay mechanisms.
Any remaining coherence in the zero-mode at non-zero temperatures
makes the condensate decay less sensitive to temperature.

\section{More than one broken symmetry}
A system may have more than one spontaneously broken symmetry.
For example, in addition to a broken gauge symmetry,
the formation of vortices breaks the rotational symmetry
of a condensate in a spherically symmetric trap \cite{vortex}.
Furthermore, rotational symmetry can also be broken \cite{ozgur}
for a multi-component or a spinor condensate \cite{mc_bec}.
When more than one continuous symmetry is broken,
there will exist as many gapless modes as for the broken symmetries,
each with its own thin spectrum. In this section,
we briefly consider the effect of more than one thin spectrum.

Consider a general effective Hamiltonian with two gapless modes
\begin{eqnarray}
\mathcal{H}=\alpha_1 p_1^2 + \alpha_2 p_2^2 + \alpha_{12} p_1 p_2+
\sum_k \hbar \omega_k b^\dagger_k b_k,
\end{eqnarray}
which after a canonical transformation, reduces to
\begin{eqnarray}
\mathcal{H}=\alpha_1' {p_1'}^2 + \alpha_2' {p_2'}^2 + \sum_k \hbar
\omega_k b^\dagger_k b_k .
\end{eqnarray}
Without loss of generality, we use this form of the Hamiltonian and
henceforth omit the primes. The observable state will be denoted by
$n$, and an easy extension leads to $\mathcal{H}|n,p_1,p_2\rangle =
E^{(p_1, p_2)}_n|n,p_1,p_2\rangle$ with $E^{(p_1,p_2)}_n =
E^{(0,0)}_n + \alpha_1 p_1^2 + \alpha_2 p_2^2$. More generally, the
primary excitation may affect both inertia terms in the two thin
spectra, which may themselves be coupled, i.e., $\alpha_1 = \alpha_1
(n,p_2)$ and $\alpha_2 = \alpha_2 (n, p_1)$. Expanding around the
small $p_1$ and $p_2$ , we find around $p_j=0$
\begin{eqnarray}
E^{(p_1 p_2)}_n
    &=& E^{(0,0)}_n + \left[\alpha_1(n,0)+\alpha'_1(n,0) p_2
        + ...\right] p_1^2 +\nonumber
        \\&&\left[\alpha_2(n,0)+\alpha'_2(n,0) p_1 + ...\right]  p_2^2\nonumber\\
    &\simeq& E^{(0,0)}_n + \alpha_1(n,0) p_1^2 + \alpha_2(n,0) p_2^2,
\end{eqnarray}
up to the second orders in $p_j$. Thus, we can safely ignore the inertia
terms' dependence on other's thin excitations to the first
approximation and let
$\alpha_1(n,p_2)=\alpha_1(n)$.
 Instead of (\ref{2statecollapse}) we now find
\begin{eqnarray}
\rho^{\rm (red)}_{\rm od}= Z^{-1} \left[\sum_{p_1 p_2} {\rm e}^{-\beta
E_0^{(p_1,p_2)}} {\rm e}^{-\frac{i}{\hbar} (E_1^{(p_1,p_2)} -E_0^{(p_1,p_2)})t}\right] C_1 C_0^*. \hskip 12pt
\end{eqnarray}
Upon substituting the approximate forms for the $E_j$s, we find
\begin{equation}
\begin{split}
\rho^{\rm (red)}_{\rm od} = & Z^{-1} {\rm e}^{-\beta
    E_0^{(0,0)}}{\rm e}^{-\frac{i}{\hbar}(E_1^{(0,0)}-E_0^{(0,0)})t} \\
        & \sum_{p_1,p_2}{\rm e}^{-\frac{i}{\hbar} \left[\alpha_1(1)-\alpha_1(0)\right]p_1^2 t}
       {\rm e}^{-\frac{i}{\hbar} \left[\alpha_2(1)-\alpha_2(0)\right]p_2^2 t} \\
       & {\rm e}^{-\beta [\alpha_1(0) p_1^2 + \alpha_2(0) p_2^2]}  C_1 C_0^* ,
\end{split}
\end{equation}

\begin{equation}
    \rho^{\rm (red)}_{\rm od}=({\rm const.}){\rm e}^{-t/t^{(1)}_{c}}{\rm e}^{-t/t^{(2)}_{c}}.
\end{equation}
Thus, we see that the collapse due to different thin spectra
do not influence each other severely. They combine to give
a resulting decay with a simple single
decay time
\begin{eqnarray}
t_{c}=\left(\frac{1}{t^{(1)}_{c}}+\frac{1}{t^{(2)}_{c}}\right)^{-1}.
\end{eqnarray}

\section{Conclusion}
Based on a toy model calculation for the decoherence dynamics of a
coherent ground state condensate, we have generalized the
calculations of the dephasing times to cases of a squeezed coherent
ground state as well as a thermal coherent ground state. The
numerical results for a squeezed ground state reveal that phase
fluctuations increase its coherence lifetime \cite{mara,ibloch},
whereas temperature increases always decrease the lifetimes for
ground state quantum coherence.

The dynamics of thin spectrum are shown to lead to decoherence, not
just on the ground state, but on quasi-particle excitations, or
superpositions of excitations. We have introduced simple
approximations that allowed for the calculations of the decoherence
lifetime of the condensate ground state as well as its coherence
excitations. These calculations make possible the discussion of
temperature effects in terms of the thermal and thermal coherent
occupations of the zero mode. We find that the lifetimes for these
two cases are of the same order of magnitude, although the lifetime
for the latter shows a weak sensitivity on temperature, whereas that
of the former displays a stronger sensitivity.

\acknowledgments

T.B. is supported by T\"UB\.ITAK. O.E.M. acknowledges the support
from a T\"UBA/GEB\.{I}P grant. L.Y. is supported by US NSF. T.B.
acknowledges a fruitful discussion with Patrick Navez.

\end{document}